\documentclass[elsarticle,reprint]{revtex4-1}
\usepackage{graphicx}
\usepackage{epstopdf}
\usepackage{epsfig}
\usepackage{amsmath}

\begin{document}


\title{Influence of magnetic field and ferromagnetic film thickness on domain pattern transfer in multiferroic heterostructures} 



\author{Diego L\'opez Gonz\'alez,$^{1}$ Arianna Casiraghi,$^{1}$ Florian Kronast,$^{2}$  K\'{e}vin J. A. Franke,$^{1}$ Sebastiaan van Dijken$^{1}$}
\email[]{sebastiaan.van.dijken@aalto.fi}
\affiliation{$^{1}$NanoSpin, Department of Applied Physics, Aalto University School of Science, P.O. Box 15100, FI-00076 Aalto, Finland.}
\affiliation{$^{2}$Helmholtz-Zentrum Berlin fuer Materialien und Energie, Albert-Einstein Str. 15, 12489 Berlin, Germany.}

\date{\today}

\begin{abstract}
We report on domain pattern transfer from a ferroelectric BaTiO$_3$ substrate to a CoFeB wedge film with a thickness of up to 150 nm. Strain coupling to domains in BaTiO$_3$ induces a regular modulation of uniaxial magnetic anisotropy in CoFeB via an inverse magnetostriction effect. As a result, the domain structures of the CoFeB wedge film and BaTiO$_3$ substrate correlate fully and straight ferroelectric domain boundaries in BaTiO$_3$ pin magnetic domain walls in CoFeB. We use x-ray photoemission electron microscopy and magneto-optical Kerr effect microscopy to characterize the spin structure of the pinned domain walls. In a rotating magnetic field, abrupt and reversible transitions between two domain wall types occur, namely, narrow walls where the magnetization vectors align head-to-tail and much broader walls with alternating head-to-head and tail-to-tail magnetization configurations. We characterize variations of the domain wall spin structure as a function of magnetic field strength and CoFeB film thickness and compare the experimental results with micromagnetic simulations.
\end{abstract}


\maketitle 

\section{Introduction}
Magnetic patterning refers to any method that induces controlled variations of magnetic properties in a continuous ferromagnetic film without topographic structuring. It allows for the creation of regular patterns of magnetic domains and domain walls, defining magnetic responses that would not occur in magnetically uniform films. Several techniques are available for magnetic patterning. These include focused ion beam irradiation \cite{CHA-98,FAS-08, FRANKEN-12, HAM-14}, low-energy proton irradiation \cite{KIM-12}, oxygen ion migration from an adjacent metal-oxide layer \cite{BAU-13,BAU-15}, and thermally-assisted scanning probe lithography \cite{ALB-16}. Interface coupling in multiferroic heterostructures can also be used to tailor the properties of ferromagnetic films. Exchange coupling to domains in a multiferroic BiFeO$_3$ layer \cite{CHU-08,LEB-09,HER-11,YOU-13,HER-14} or strain coupling to ferroelastic domains in a BaTiO$_3$ crystal \cite{LAH-11,LAH-12,LAH-12a,CHO-12,STR-13,GHI-15,FRA-15,SHI-15}, for instance, induce lateral modulations of magnetic anisotropy in the adjacent ferromagnetic film. If the modulations of magnetic anisotropy are strong enough, one-to-one correlations between domain patterns in the ferroelectric and ferromagnetic layers are attained. Full domain pattern transfer in multiferroic heterostructures enables electric-field induced magnetic switching \cite{CHU-08,LEB-09,HER-11,HER-14,LAH-11,LAH-12,GHI-15,SHI-15} and reversible magnetic domain wall motion \cite{LAH-12,FRA-15}, offering promising prospects for low-power spintronic devices.  

\begin{figure}[t!]
\includegraphics{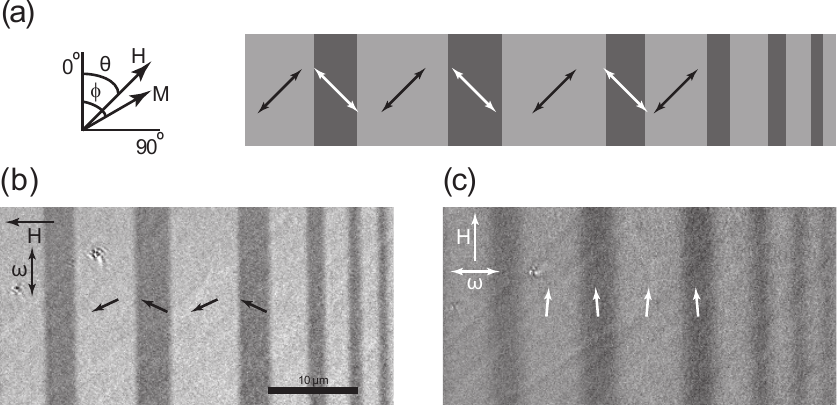}
\caption{\label{Fig1}(a) Schematic illustration of magnetic stripe domains in the CoFeB film. The double headed arrows indicate the modulation of uniaxial magnetic anisotropy. The angle of in-plane magnetic field ($\theta$) and the direction of magnetization within stripe domains ($\phi$) are defined on the left. (b),(c) MOKE microscopy images of the CoFeB film with uncharged and charged magnetic domain walls, respectively. The strength of applied magnetic field is 40 mT and the CoFeB film is 150 nm thick. The sensitive axis of MOKE contrast is indicated by $\omega$.}
\end{figure}

Here, we investigate how domain pattern transfer from a ferroelectric BaTiO$_3$ substrate to a CoFeB wedge film varies with ferromagnetic film thickness and applied magnetic field. In our experiments, the BaTiO$_3$ substrate consists of stripe domains with in-plane ferroelectric polarization. At the domain boundaries, the polarization rotates abruptly by 90$^{\circ}$. Since the BaTiO$_3$ crystal is tetragonal at room temperature, rotation of the ferroelectric polarization also reorients the axis of lattice tetragonality. Via inverse magnetostriction, this induces regular modulations of uniaxial magnetic anisotropy in the CoFeB wedge film, as illustrated in Fig. \ref{Fig1}(a). Consequently, the ferroelectric stripe domain pattern in the BaTiO$_3$ substrate is fully transferred to the CoFeB wedge film and the strain-induced rotations of uniaxial magnetic anisotropy at the ferroelectric domain boundaries firmly pin the magnetic domain walls. Magnetic domain walls are therefore as straight as their ferroelectric counterpart and do not move in a magnetic field. The latter effect enables active switching between two types of magnetic domain walls \cite{FRA-12}. Domain walls with a head-to-tail alignment of magnetization form when an in-plane magnetic field is applied perpendicular to the stripe domains (Fig. \ref{Fig1}(b)). The energy and width of these walls are determined mostly by the strength of exchange interactions and magnetic anisotropy. An in-plane magnetic field parallel to the stripe domains, on the other hand, initializes domain walls with alternating head-to-head and tail-to-tail magnetization configurations (Fig. \ref{Fig1}(c)). Being defined by magnetic anisotropy and long-range magnetostatic interactions (i.e. magnetic charges), these magnetic domain walls are much wider and higher in energy than the head-to-tail walls. Hereafter, we will refer to the two domain-wall types as magnetically charged (head-to-head and tail-to-tail) and uncharged (head-to-tail). 

\begin{figure}[t!]
\includegraphics{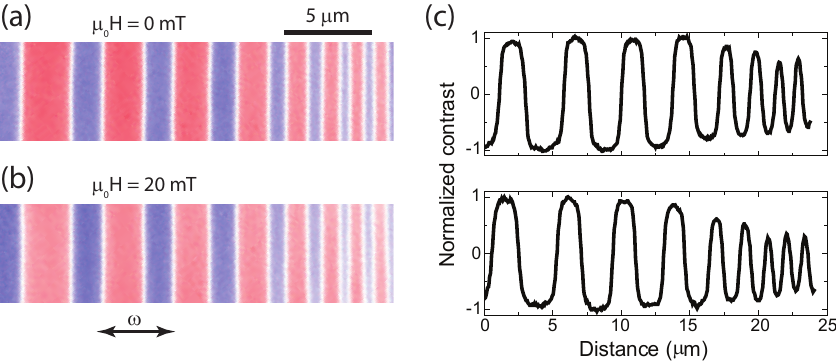}
\caption{\label{Fig2}(a),(b) PEEM images of the CoFeB film with charged domain walls. The images in (a) and (b) are recorded in zero magnetic field and a magnetic field of 20 mT applied at $\theta=0^{\circ}$, respectively. The CoFeB film is 150 nm thick. (c) Line profiles of the normalized XMCD contrast in (a) and (b).}
\end{figure}

The width and energy of pinned magnetic domain walls affects domain pattern transfer and magnetic switching in multiferroic heterostructures \cite{FRA-14,CAS-15}. Breakdown of pattern transfer occurs when the width of magnetic domain walls and ferroelectric stripe domains become comparable. More specifically, in the strain-coupled CoFeB/BaTiO$_3$ system we consider here, the rotation of magnetization between neighboring domains in zero magnetic field decreases from 90$^{\circ}$ for wide ferroelectric stripe domains to nearly 0$^{\circ}$ for very narrow domains. As the widths of charged and uncharged magnetic domain walls differ significantly (up to about two orders of magnitude in thick ferromagnetic films), magnetic switching from one domain wall type to the other changes the amount of magnetization rotation between domains and, thereby, the magnetic contrast. An example is shown in Figs. \ref{Fig1}(b) and (c). The magneto-optical Kerr effect (MOKE) microscopy images in these figures are recorded on a sample area where the CoFeB wedge film is 150 nm thick. In (b), a 40 mT in-plane magnetic field is oriented perpendicular to the stripe domains. The uncharged magnetic domain walls that form are much narrower than the stripe domains and, consequently, a sizeable rotation of magnetization between domains and clear magnetic contrast are obtained. In (c), the in-plane magnetic field is rotated by 90$^{\circ}$. Now, much wider charged domain walls separate the stripe domains and the magnetization in the CoFeB film is more uniform, particularly for the narrow stripe domains on the right. Rotation of an external magnetic field thus causes the consecutive writing and erasure of magnetic stripe patterns, as previously discussed in Ref. \citenum{FRA-14}. In this paper, we report on the variation of domain pattern transfer with ferromagnetic film thickness and the strength and orientation of applied magnetic field. Experimentally, we measure the evolution of magnetic domain patterns in a rotating magnetic field. From these measurements we derive $|\phi_1 - \phi_2|$, where $\phi_1$ and $\phi_2$ indicate the angles of magnetization in two neighboring stripe domains. Large values of $|\phi_1 - \phi_2|$ thus represent robust domain pattern transfer. We use micromagnetic simulations to interpret our experimental data.             

\section{Methods}
For the experiments we grew a Co$_{40}$Fe$_{40}$B$_{20}$ wedge film, with thickness $t$ up to 150 nm, onto a BaTiO$_3$ substrate. The film was sputtered at 300 $^{\circ}$C and capped with 3 nm Au to prevent oxidation. At this deposition temperature the BaTiO$_3$ substrate is paraelectric and its crystal structure is cubic. The ferroelectric and ferromagnetic stripe domains are formed during post-deposition cooling through the BaTiO$_3$ paraelectric-to-ferroelectric phase transition at $T_C\approx120^\circ$C. This procedure resulted in one-to-one domain correlations across the entire CoFeB wedge film. We used MOKE microscopy to image the magnetic stripe domains in a slowly rotating in-plane magnetic field. High-resolution images of the magnetic domains were obtained by x-ray photoemission electron microscopy (PEEM) on beamline UE49-PGM1 at the BESSY II synchrotron facility. Illumination at oblique incidence and x-ray magnetic circular dichroism (XMCD) at the Fe L$_3$ edge were used to visualize the direction of in-plane magnetization with a spatial resolution of about 30 nm. Micromagnetic simulations were performed in MuMax3 \cite{VAN-14}. In the simulations, the width of the stripe domains was set to 3 $\mu$m and the simulation area was discretized into $3\times3\times{t}$ nm$^3$ cells. As input parameters we used a uniaxial magnetic anisotropy $K_u=3\times10^4$ J/m$^3$ (derived from MOKE measurements), a saturation magnetization $M_s=1.2\times10^6$ A/m, and an exchange constant $A=2.1\times10^{-11}$ J/m.  

\section{Results and discussion}
Figure \ref{Fig2} shows PEEM images of the CoFeB film (a) at remanence and (b) under an in-plane magnetic field of 20 mT. In these experiments, charged domain walls separate the magnetic stripe domains and the CoFeB film is 150 nm thick. From line profiles on wide domains (Fig. \ref{Fig2}(c)), we extract a domain wall width of 2.0 $\mu$m for $\mu_0H=0$ mT and 2.1 $\mu$m for $\mu_0H=20$ mT. Because of the large width of charged domain walls, the magnetic contrast reduces considerably in the narrow stripe domains on the right side of the PEEM images. This finite-size scaling effect is slightly more pronounced for $\mu_0H=20$ mT, which confirms the notion that wider domain walls cause a stronger reduction of magnetization rotation in narrow domains.   

\begin{figure}[t!]
\includegraphics{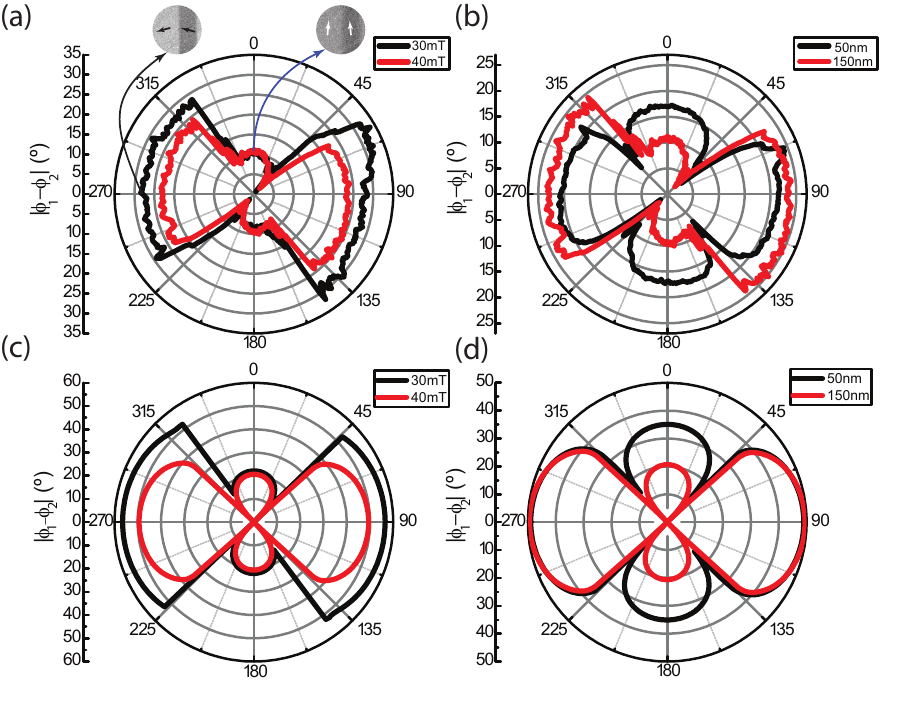}
\caption{\label{Fig3} Magnetization rotation between two neighboring stripe domains as a function of magnetic field angle. In (a), the CoFeB film is 150 nm thick and results for a magnetic field of 30 mT and 40 mT are shown. In (b), the magnetic field is 40 mT and results for a CoFeB film thickness of 50 nm and 150 nm are presented. Panels (c) and (d) show results from micromagnetic simulations.}
\end{figure}

Next, we investigate the effect of a rotating in-plane magnetic field on domain pattern transfer. In the experiments, MOKE microscopy images are recorded as a function of magnetic field angle $\theta$. From the series of images we extract the orientation of magnetization ($\phi_1$ and $\phi_2$) in two neighboring stripe domains. The stripe domains are 3 - 4 $\mu$m wide. Figures \ref{Fig3}(a) and (b) show results for a rotating magnetic field of 30 mT and 40 mT and a CoFeB film thickness of 50 nm and 150 nm. In all cases, transitions between small and larger values of $|\phi_1 - \phi_2|$ are measured. These switching events, which signify a change in magnetic contrast (see Fig. \ref{Fig1}), are caused by abrupt domain wall transformations. Starting from $\theta=0^{\circ}$, the stripe domains are initially separated by broad charged domain walls. The rotation of magnetization between neighboring domains is thus small. Clockwise rotation of a constant magnetic field switches the magnetization of every second stripe domain at $\theta\approx45-55^{\circ}$. As a result, the broad charged domain walls transform into narrow uncharged domain walls. This increases $|\phi_1 - \phi_2|$ and, consequently, the magnetic contrast of MOKE microscopy images. At $\theta\approx135-145^{\circ}$, a second switching event restores the charged domain walls, causing a reduction of $|\phi_1 - \phi_2|$. Back and forth switching between the two types of domain walls continues upon further rotation of the magnetic field. For narrow uncharged domain walls, the ferroelectric stripe domain pattern in the BaTiO$_3$ substrate is clearly transferred to the CoFeB wedge. Broad charged domain walls, on the other hand, lead to nearly uniform magnetization in the CoFeB film. Results from micromagnetic simulations, shown in Figs. \ref{Fig3}(c) and (d), qualitatively confirm the measurements. While the shape of the switching curves is nearly identical, the magnitude of $|\phi_1 - \phi_2|$ is larger in the simulations. The main reason for this is a difference in data acquisition. In order to increase the signal to noise ratio in the experimental data, we extract the average direction of magnetization in two neighboring stripe domains, while simulated data is collected at the two domain centers.             

\begin{figure}[t!]
\includegraphics{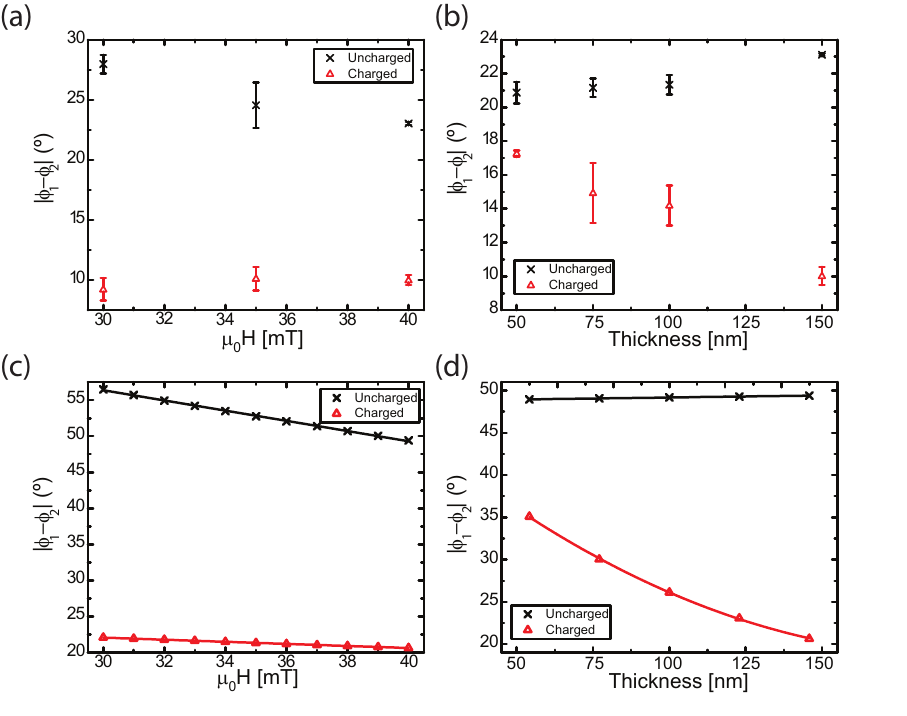}
\caption{\label{Fig4}  Magnetization rotation between two neighboring stripe domains for uncharged and charged magnetic domain walls. The data are acquired for $\theta=0^{\circ}$ and $\theta=90^{\circ}$. In (a), the dependence on magnetic field is plotted for a CoFeB film thickness of 150 nm. The influence of CoFeB film thickness is shown in (b) for $\mu_0H=40$ mT. Corresponding results from micromagnetic simulations are depicted in (c) and (d).}
\end{figure}

Figure \ref{Fig4} summarizes the dependence of $|\phi_1 - \phi_2|$ on applied magnetic field and CoFeB film thickness. With increasing field, the magnetization rotates towards the field axis. The resulting reduction in the rotation of magnetization between neighboring stripe domains is more pronounced for uncharged domain walls. The effect of CoFeB film thickness is opposite for the two types of magnetic domain walls. The width of charged domain walls increases with ferromagnetic film thickness due to stronger magnetostatic interactions. For $180^{\circ}$ charged walls, the width is given by $\delta_{c}=2\pi\mu_{0}M_{s}^{2}t/8K_{u}$ \cite{HUB-79}. Although the exact profile of charged domain walls changes upon a reduction of magnetization rotation, the width of smaller-angle walls does also increase with ferromagnetic film thickness. Wider charged domain walls on the thick side of the CoFeB wedge film significantly reduce $|\phi_1 - \phi_2|$ and, thus, the degree of domain pattern transfer. We measure and simulate a much smaller and opposite effect for uncharged domain walls. While the spin structure of these narrow walls is mostly determined by the strength of exchange interactions and uniaxial magnetic anisotropy, magnetostatic interactions slightly reduce their width in thick films \cite{FRA-14}. This effect enhances the rotation of magnetization between neighboring domains, in agreement with the data of Figs. \ref{Fig4}(b) and (d).   

\begin{figure}[t!]
\includegraphics{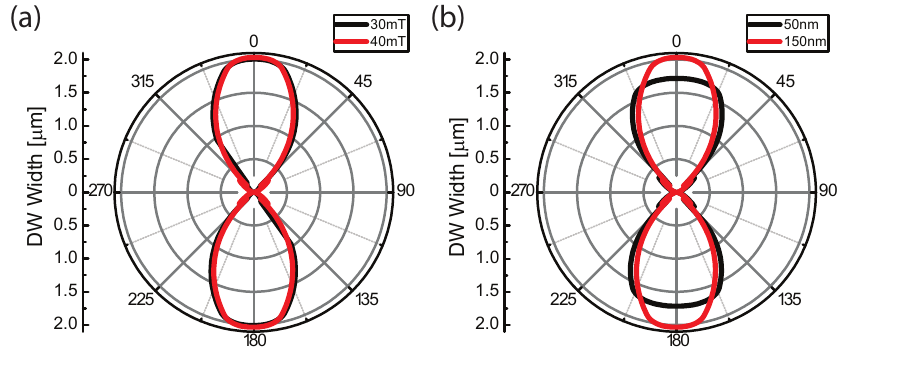}
\caption{\label{Fig5} Simulated domain wall width as a function of magnetic field angle for a 150 nm thick CoFeB film (a) and a magnetic field of 40 mT (b).}
\end{figure}

Finally, we use micromagnetic simulations to estimate how the width of pinned magnetic domain walls in our CoFeB/BaTiO$_3$ multiferroic heterostructure varies as a function of magnetic field angle. We define the domain wall width as \cite{FRA-14}
  
\begin{equation}
\delta=\int_{-\infty}^{+\infty}cos^2 (\phi') dx,
\end{equation}
where $\phi'$ is the reduced magnetization angle 

\begin{equation}
\phi' = \left(\phi(x)-\frac{|\phi_1-\phi_2|}{2}\right)\frac{180}{|\phi_1-\phi_2|},
\end{equation}
and $\phi_1$ and $\phi_2$ are the angles of magnetization in the centers of two neighboring stripe domains. For zero magnetic field and a CoFeB film thickness of 150 nm, we simulate a width of 25 nm and 1.9 $\mu$m for uncharged and charged domain walls, respectively, a difference of nearly two orders of magnitude. The width of charged magnetic domain walls agrees well with the PEEM measurements of Fig. \ref{Fig2}. 
   
Figure \ref{Fig5} shows the variation of domain wall width in a rotating in-plane magnetic field. The widest domain walls are obtained if the magnetic field is aligned parallel to the stripe domains. Rotation of the magnetic field away from this position reduces the width of these charged domain walls. At $\theta=45^{\circ}$, a transformation to much narrower uncharged domain walls occurs. Back and forth switching between the two types of domain walls in a rotating magnetic field changes the wall width by about two orders of magnitude for $t=150$ nm. Thus, while the walls remain firmly pinned onto the ferroelectric boundaries in the BaTiO$_3$ substrate, the magnetic field tunes their spin structure and width. As a result, clear imprinting of narrow BaTiO$_3$ stripe domains into the CoFeB film occurs only if the magnetic domain walls are uncharged.   

In conclusions, we have studied domain pattern transfer from a ferroelectric BaTiO$_3$ substrate to a ferromagnetic CoFeB wedge film using MOKE microscopy, x-ray PEEM, and micromagnetic simulations. This strain-coupled multiferroic system is characterized by strong pinning of magnetic domain walls onto ferroelectric boundaries. Using a rotating magnetic field, we have demonstrated that the width of pinned magnetic domain walls can be tuned by two orders of magnitude. Switching from narrow uncharged domain walls to broad charged domain walls deteriorates the transfer of domain patterns, which is reflected by a decrease of magnetic contrast in MOKE microscopy images. The weakest magnetic contrast is obtained when the stripe domains are narrow and the CoFeB film is thick. \\

\section{Acknowledgments}
This work was supported by the European Research Council (ERC-2012-StG 307502-E-CONTROL and ERC-2014-PoC 665215-EMOTION).


\begin{thebibliography}{26}%
\makeatletter
\providecommand \@ifxundefined [1]{%
 \@ifx{#1\undefined}
}%
\providecommand \@ifnum [1]{%
 \ifnum #1\expandafter \@firstoftwo
 \else \expandafter \@secondoftwo
 \fi
}%
\providecommand \@ifx [1]{%
 \ifx #1\expandafter \@firstoftwo
 \else \expandafter \@secondoftwo
 \fi
}%
\providecommand \natexlab [1]{#1}%
\providecommand \enquote  [1]{``#1''}%
\providecommand \bibnamefont  [1]{#1}%
\providecommand \bibfnamefont [1]{#1}%
\providecommand \citenamefont [1]{#1}%
\providecommand \href@noop [0]{\@secondoftwo}%
\providecommand \href [0]{\begingroup \@sanitize@url \@href}%
\providecommand \@href[1]{\@@startlink{#1}\@@href}%
\providecommand \@@href[1]{\endgroup#1\@@endlink}%
\providecommand \@sanitize@url [0]{\catcode `\\12\catcode `\$12\catcode
  `\&12\catcode `\#12\catcode `\^12\catcode `\_12\catcode `\%12\relax}%
\providecommand \@@startlink[1]{}%
\providecommand \@@endlink[0]{}%
\providecommand \url  [0]{\begingroup\@sanitize@url \@url }%
\providecommand \@url [1]{\endgroup\@href {#1}{\urlprefix }}%
\providecommand \urlprefix  [0]{URL }%
\providecommand \Eprint [0]{\href }%
\providecommand \doibase [0]{http://dx.doi.org/}%
\providecommand \selectlanguage [0]{\@gobble}%
\providecommand \bibinfo  [0]{\@secondoftwo}%
\providecommand \bibfield  [0]{\@secondoftwo}%
\providecommand \translation [1]{[#1]}%
\providecommand \BibitemOpen [0]{}%
\providecommand \bibitemStop [0]{}%
\providecommand \bibitemNoStop [0]{.\EOS\space}%
\providecommand \EOS [0]{\spacefactor3000\relax}%
\providecommand \BibitemShut  [1]{\csname bibitem#1\endcsname}%
\let\auto@bib@innerbib\@empty
\bibitem [{\citenamefont {{Chappert}}\ \emph {et~al.}(1998)\citenamefont
  {{Chappert}}, \citenamefont {{Bernas}}, \citenamefont {{Ferre}},
  \citenamefont {{Kottler}}, \citenamefont {{Jamet}}, \citenamefont {{Chen}},
  \citenamefont {{Cambril}}, \citenamefont {{Devolder}}, \citenamefont
  {{Rousseaux}}, \citenamefont {{Mathet}},\ and\ \citenamefont
  {{Launois}}}]{CHA-98}%
  \BibitemOpen
  \bibfield  {author} {\bibinfo {author} {\bibfnamefont {C.}~\bibnamefont
  {{Chappert}}}, \bibinfo {author} {\bibfnamefont {H.}~\bibnamefont
  {{Bernas}}}, \bibinfo {author} {\bibfnamefont {J.}~\bibnamefont {{Ferre}}},
  \bibinfo {author} {\bibfnamefont {V.}~\bibnamefont {{Kottler}}}, \bibinfo
  {author} {\bibfnamefont {J.-P.}\ \bibnamefont {{Jamet}}}, \bibinfo {author}
  {\bibfnamefont {Y.}~\bibnamefont {{Chen}}}, \bibinfo {author} {\bibfnamefont
  {E.}~\bibnamefont {{Cambril}}}, \bibinfo {author} {\bibfnamefont
  {T.}~\bibnamefont {{Devolder}}}, \bibinfo {author} {\bibfnamefont
  {F.}~\bibnamefont {{Rousseaux}}}, \bibinfo {author} {\bibfnamefont
  {V.}~\bibnamefont {{Mathet}}}, \ and\ \bibinfo {author} {\bibfnamefont
  {H.}~\bibnamefont {{Launois}}},\ }\href {\doibase
  10.1126/science.280.5371.1919} {\bibfield  {journal} {\bibinfo  {journal}
  {Science}\ }\textbf {\bibinfo {volume} {280}},\ \bibinfo {pages} {1919}
  (\bibinfo {year} {1998})}\BibitemShut {NoStop}%
\bibitem [{\citenamefont {{Fassbender}}\ and\ \citenamefont
  {{McCord}}(2008)}]{FAS-08}%
  \BibitemOpen
  \bibfield  {author} {\bibinfo {author} {\bibfnamefont {J.}~\bibnamefont
  {{Fassbender}}}\ and\ \bibinfo {author} {\bibfnamefont {J.}~\bibnamefont
  {{McCord}}},\ }\href {\doibase 10.1016/j.jmmm.2007.07.032} {\bibfield
  {journal} {\bibinfo  {journal} {Journal of Magnetism and Magnetic Materials}\
  }\textbf {\bibinfo {volume} {320}},\ \bibinfo {pages} {579} (\bibinfo {year}
  {2008})}\BibitemShut {NoStop}%
\bibitem [{\citenamefont {{Franken}}\ \emph {et~al.}(2012)\citenamefont
  {{Franken}}, \citenamefont {{Swagten}},\ and\ \citenamefont
  {{Koopmans}}}]{FRANKEN-12}%
  \BibitemOpen
  \bibfield  {author} {\bibinfo {author} {\bibfnamefont {J.~H.}\ \bibnamefont
  {{Franken}}}, \bibinfo {author} {\bibfnamefont {H.~J.~M.}\ \bibnamefont
  {{Swagten}}}, \ and\ \bibinfo {author} {\bibfnamefont {B.}~\bibnamefont
  {{Koopmans}}},\ }\href {\doibase 10.1038/nnano.2012.111} {\bibfield
  {journal} {\bibinfo  {journal} {Nature Nanotechnology}\ }\textbf {\bibinfo
  {volume} {7}},\ \bibinfo {pages} {499} (\bibinfo {year} {2012})}\BibitemShut
  {NoStop}%
\bibitem [{\citenamefont {{Hamann}}\ \emph {et~al.}(2014)\citenamefont
  {{Hamann}}, \citenamefont {{Mattheis}}, \citenamefont {{M{\"o}nch}},
  \citenamefont {{Fassbender}}, \citenamefont {{Schultz}},\ and\ \citenamefont
  {{McCord}}}]{HAM-14}%
  \BibitemOpen
  \bibfield  {author} {\bibinfo {author} {\bibfnamefont {C.}~\bibnamefont
  {{Hamann}}}, \bibinfo {author} {\bibfnamefont {R.}~\bibnamefont
  {{Mattheis}}}, \bibinfo {author} {\bibfnamefont {I.}~\bibnamefont
  {{M{\"o}nch}}}, \bibinfo {author} {\bibfnamefont {J.}~\bibnamefont
  {{Fassbender}}}, \bibinfo {author} {\bibfnamefont {L.}~\bibnamefont
  {{Schultz}}}, \ and\ \bibinfo {author} {\bibfnamefont {J.}~\bibnamefont
  {{McCord}}},\ }\href {\doibase 10.1088/1367-2630/16/2/023010} {\bibfield
  {journal} {\bibinfo  {journal} {New Journal of Physics}\ }\textbf {\bibinfo
  {volume} {16}},\ \bibinfo {pages} {023010} (\bibinfo {year}
  {2014})}\BibitemShut {NoStop}%
\bibitem [{\citenamefont {{Kim}}\ \emph {et~al.}(2012)\citenamefont {{Kim}},
  \citenamefont {{Lee}}, \citenamefont {{Ko}}, \citenamefont {{Son}},
  \citenamefont {{Kim}}, \citenamefont {{Kang}},\ and\ \citenamefont
  {{Hong}}}]{KIM-12}%
  \BibitemOpen
  \bibfield  {author} {\bibinfo {author} {\bibfnamefont {S.}~\bibnamefont
  {{Kim}}}, \bibinfo {author} {\bibfnamefont {S.}~\bibnamefont {{Lee}}},
  \bibinfo {author} {\bibfnamefont {J.}~\bibnamefont {{Ko}}}, \bibinfo {author}
  {\bibfnamefont {J.}~\bibnamefont {{Son}}}, \bibinfo {author} {\bibfnamefont
  {M.}~\bibnamefont {{Kim}}}, \bibinfo {author} {\bibfnamefont
  {S.}~\bibnamefont {{Kang}}}, \ and\ \bibinfo {author} {\bibfnamefont
  {J.}~\bibnamefont {{Hong}}},\ }\href {\doibase 10.1038/nnano.2012.125}
  {\bibfield  {journal} {\bibinfo  {journal} {Nature Nanotechnology}\ }\textbf
  {\bibinfo {volume} {7}},\ \bibinfo {pages} {567} (\bibinfo {year}
  {2012})}\BibitemShut {NoStop}%
\bibitem [{\citenamefont {{Bauer}}\ \emph {et~al.}(2013)\citenamefont
  {{Bauer}}, \citenamefont {{Emori}},\ and\ \citenamefont {{Beach}}}]{BAU-13}%
  \BibitemOpen
  \bibfield  {author} {\bibinfo {author} {\bibfnamefont {U.}~\bibnamefont
  {{Bauer}}}, \bibinfo {author} {\bibfnamefont {S.}~\bibnamefont {{Emori}}}, \
  and\ \bibinfo {author} {\bibfnamefont {G.~S.~D.}\ \bibnamefont {{Beach}}},\
  }\href {\doibase 10.1038/nnano.2013.96} {\bibfield  {journal} {\bibinfo
  {journal} {Nature Nanotechnology}\ }\textbf {\bibinfo {volume} {8}},\
  \bibinfo {pages} {411} (\bibinfo {year} {2013})}\BibitemShut {NoStop}%
\bibitem [{\citenamefont {{Bauer}}\ \emph {et~al.}(2015)\citenamefont
  {{Bauer}}, \citenamefont {{Yao}}, \citenamefont {{Tan}}, \citenamefont
  {{Agrawal}}, \citenamefont {{Emori}}, \citenamefont {{Tuller}}, \citenamefont
  {{van Dijken}},\ and\ \citenamefont {{Beach}}}]{BAU-15}%
  \BibitemOpen
  \bibfield  {author} {\bibinfo {author} {\bibfnamefont {U.}~\bibnamefont
  {{Bauer}}}, \bibinfo {author} {\bibfnamefont {L.}~\bibnamefont {{Yao}}},
  \bibinfo {author} {\bibfnamefont {A.~J.}\ \bibnamefont {{Tan}}}, \bibinfo
  {author} {\bibfnamefont {P.}~\bibnamefont {{Agrawal}}}, \bibinfo {author}
  {\bibfnamefont {S.}~\bibnamefont {{Emori}}}, \bibinfo {author} {\bibfnamefont
  {H.~L.}\ \bibnamefont {{Tuller}}}, \bibinfo {author} {\bibfnamefont
  {S.}~\bibnamefont {{van Dijken}}}, \ and\ \bibinfo {author} {\bibfnamefont
  {G.~S.~D.}\ \bibnamefont {{Beach}}},\ }\href {\doibase 10.1038/nmat4134}
  {\bibfield  {journal} {\bibinfo  {journal} {Nature Materials}\ }\textbf
  {\bibinfo {volume} {14}},\ \bibinfo {pages} {174} (\bibinfo {year}
  {2015})}\BibitemShut {NoStop}%
\bibitem [{\citenamefont {{Albisetti}}\ \emph {et~al.}(2016)\citenamefont
  {{Albisetti}}, \citenamefont {{Petti}}, \citenamefont {{Pancaldi}},
  \citenamefont {{Madami}}, \citenamefont {{Tacchi}}, \citenamefont {{Curtis}},
  \citenamefont {{King}}, \citenamefont {{Papp}}, \citenamefont {{Csaba}},
  \citenamefont {{Porod}}, \citenamefont {{Vavassori}}, \citenamefont
  {{Riedo}},\ and\ \citenamefont {{Bertacco}}}]{ALB-16}%
  \BibitemOpen
  \bibfield  {author} {\bibinfo {author} {\bibfnamefont {E.}~\bibnamefont
  {{Albisetti}}}, \bibinfo {author} {\bibfnamefont {D.}~\bibnamefont
  {{Petti}}}, \bibinfo {author} {\bibfnamefont {M.}~\bibnamefont {{Pancaldi}}},
  \bibinfo {author} {\bibfnamefont {M.}~\bibnamefont {{Madami}}}, \bibinfo
  {author} {\bibfnamefont {S.}~\bibnamefont {{Tacchi}}}, \bibinfo {author}
  {\bibfnamefont {J.}~\bibnamefont {{Curtis}}}, \bibinfo {author}
  {\bibfnamefont {W.~P.}\ \bibnamefont {{King}}}, \bibinfo {author}
  {\bibfnamefont {A.}~\bibnamefont {{Papp}}}, \bibinfo {author} {\bibfnamefont
  {G.}~\bibnamefont {{Csaba}}}, \bibinfo {author} {\bibfnamefont
  {W.}~\bibnamefont {{Porod}}}, \bibinfo {author} {\bibfnamefont
  {P.}~\bibnamefont {{Vavassori}}}, \bibinfo {author} {\bibfnamefont
  {E.}~\bibnamefont {{Riedo}}}, \ and\ \bibinfo {author} {\bibfnamefont
  {R.}~\bibnamefont {{Bertacco}}},\ }\href {\doibase 10.1038/nnano.2016.25}
  {\bibfield  {journal} {\bibinfo  {journal} {Nature Nanotechnology}\ }\textbf
  {\bibinfo {volume} {11}},\ \bibinfo {pages} {545} (\bibinfo {year}
  {2016})}\BibitemShut {NoStop}%
\bibitem [{\citenamefont {{Chu}}\ \emph {et~al.}(2008)\citenamefont {{Chu}},
  \citenamefont {{Martin}}, \citenamefont {{Holcomb}}, \citenamefont {{Gajek}},
  \citenamefont {{Han}}, \citenamefont {{He}}, \citenamefont {{Balke}},
  \citenamefont {{Yang}}, \citenamefont {{Lee}}, \citenamefont {{Hu}},
  \citenamefont {{Zhan}}, \citenamefont {{Yang}}, \citenamefont
  {{Fraile-Rodr{\'{\i}}guez}}, \citenamefont {{Scholl}}, \citenamefont
  {{Wang}},\ and\ \citenamefont {{Ramesh}}}]{CHU-08}%
  \BibitemOpen
  \bibfield  {author} {\bibinfo {author} {\bibfnamefont {Y.-H.}\ \bibnamefont
  {{Chu}}}, \bibinfo {author} {\bibfnamefont {L.~W.}\ \bibnamefont {{Martin}}},
  \bibinfo {author} {\bibfnamefont {M.~B.}\ \bibnamefont {{Holcomb}}}, \bibinfo
  {author} {\bibfnamefont {M.}~\bibnamefont {{Gajek}}}, \bibinfo {author}
  {\bibfnamefont {S.-J.}\ \bibnamefont {{Han}}}, \bibinfo {author}
  {\bibfnamefont {Q.}~\bibnamefont {{He}}}, \bibinfo {author} {\bibfnamefont
  {N.}~\bibnamefont {{Balke}}}, \bibinfo {author} {\bibfnamefont {C.-H.}\
  \bibnamefont {{Yang}}}, \bibinfo {author} {\bibfnamefont {D.}~\bibnamefont
  {{Lee}}}, \bibinfo {author} {\bibfnamefont {W.}~\bibnamefont {{Hu}}},
  \bibinfo {author} {\bibfnamefont {Q.}~\bibnamefont {{Zhan}}}, \bibinfo
  {author} {\bibfnamefont {P.-L.}\ \bibnamefont {{Yang}}}, \bibinfo {author}
  {\bibfnamefont {A.}~\bibnamefont {{Fraile-Rodr{\'{\i}}guez}}}, \bibinfo
  {author} {\bibfnamefont {A.}~\bibnamefont {{Scholl}}}, \bibinfo {author}
  {\bibfnamefont {S.~X.}\ \bibnamefont {{Wang}}}, \ and\ \bibinfo {author}
  {\bibfnamefont {R.}~\bibnamefont {{Ramesh}}},\ }\href {\doibase
  10.1038/nmat2184} {\bibfield  {journal} {\bibinfo  {journal} {Nature
  Materials}\ }\textbf {\bibinfo {volume} {7}},\ \bibinfo {pages} {478}
  (\bibinfo {year} {2008})}\BibitemShut {NoStop}%
\bibitem [{\citenamefont {{Lebeugle}}\ \emph {et~al.}(2009)\citenamefont
  {{Lebeugle}}, \citenamefont {{Mougin}}, \citenamefont {{Viret}},
  \citenamefont {{Colson}},\ and\ \citenamefont {{Ranno}}}]{LEB-09}%
  \BibitemOpen
  \bibfield  {author} {\bibinfo {author} {\bibfnamefont {D.}~\bibnamefont
  {{Lebeugle}}}, \bibinfo {author} {\bibfnamefont {A.}~\bibnamefont
  {{Mougin}}}, \bibinfo {author} {\bibfnamefont {M.}~\bibnamefont {{Viret}}},
  \bibinfo {author} {\bibfnamefont {D.}~\bibnamefont {{Colson}}}, \ and\
  \bibinfo {author} {\bibfnamefont {L.}~\bibnamefont {{Ranno}}},\ }\href
  {\doibase 10.1103/PhysRevLett.103.257601} {\bibfield  {journal} {\bibinfo
  {journal} {Physical Review Letters}\ }\textbf {\bibinfo {volume} {103}},\
  \bibinfo {pages} {257601} (\bibinfo {year} {2009})}\BibitemShut {NoStop}%
\bibitem [{\citenamefont {{Heron}}\ \emph {et~al.}(2011)\citenamefont
  {{Heron}}, \citenamefont {{Trassin}}, \citenamefont {{Ashraf}}, \citenamefont
  {{Gajek}}, \citenamefont {{He}}, \citenamefont {{Yang}}, \citenamefont
  {{Nikonov}}, \citenamefont {{Chu}}, \citenamefont {{Salahuddin}},\ and\
  \citenamefont {{Ramesh}}}]{HER-11}%
  \BibitemOpen
  \bibfield  {author} {\bibinfo {author} {\bibfnamefont {J.~T.}\ \bibnamefont
  {{Heron}}}, \bibinfo {author} {\bibfnamefont {M.}~\bibnamefont {{Trassin}}},
  \bibinfo {author} {\bibfnamefont {K.}~\bibnamefont {{Ashraf}}}, \bibinfo
  {author} {\bibfnamefont {M.}~\bibnamefont {{Gajek}}}, \bibinfo {author}
  {\bibfnamefont {Q.}~\bibnamefont {{He}}}, \bibinfo {author} {\bibfnamefont
  {S.~Y.}\ \bibnamefont {{Yang}}}, \bibinfo {author} {\bibfnamefont {D.~E.}\
  \bibnamefont {{Nikonov}}}, \bibinfo {author} {\bibfnamefont {Y.-H.}\
  \bibnamefont {{Chu}}}, \bibinfo {author} {\bibfnamefont {S.}~\bibnamefont
  {{Salahuddin}}}, \ and\ \bibinfo {author} {\bibfnamefont {R.}~\bibnamefont
  {{Ramesh}}},\ }\href {\doibase 10.1103/PhysRevLett.107.217202} {\bibfield
  {journal} {\bibinfo  {journal} {Physical Review Letters}\ }\textbf {\bibinfo
  {volume} {107}},\ \bibinfo {pages} {217202} (\bibinfo {year}
  {2011})}\BibitemShut {NoStop}%
\bibitem [{\citenamefont {{You}}\ \emph {et~al.}(2013)\citenamefont {{You}},
  \citenamefont {{Wang}}, \citenamefont {{Zou}}, \citenamefont {{Lim}},
  \citenamefont {{Zhou}}, \citenamefont {{Ding}}, \citenamefont {{Chen}},\ and\
  \citenamefont {{Wang}}}]{YOU-13}%
  \BibitemOpen
  \bibfield  {author} {\bibinfo {author} {\bibfnamefont {L.}~\bibnamefont
  {{You}}}, \bibinfo {author} {\bibfnamefont {B.}~\bibnamefont {{Wang}}},
  \bibinfo {author} {\bibfnamefont {X.}~\bibnamefont {{Zou}}}, \bibinfo
  {author} {\bibfnamefont {Z.~S.}\ \bibnamefont {{Lim}}}, \bibinfo {author}
  {\bibfnamefont {Y.}~\bibnamefont {{Zhou}}}, \bibinfo {author} {\bibfnamefont
  {H.}~\bibnamefont {{Ding}}}, \bibinfo {author} {\bibfnamefont
  {L.}~\bibnamefont {{Chen}}}, \ and\ \bibinfo {author} {\bibfnamefont
  {J.}~\bibnamefont {{Wang}}},\ }\href {\doibase 10.1103/PhysRevB.88.184426}
  {\bibfield  {journal} {\bibinfo  {journal} {Physical Review B}\ }\textbf
  {\bibinfo {volume} {88}},\ \bibinfo {pages} {184426} (\bibinfo {year}
  {2013})}\BibitemShut {NoStop}%
\bibitem [{\citenamefont {{Heron}}\ \emph {et~al.}(2014)\citenamefont
  {{Heron}}, \citenamefont {{Bosse}}, \citenamefont {{He}}, \citenamefont
  {{Gao}}, \citenamefont {{Trassin}}, \citenamefont {{Ye}}, \citenamefont
  {{Clarkson}}, \citenamefont {{Wang}}, \citenamefont {{Liu}}, \citenamefont
  {{Salahuddin}}, \citenamefont {{Ralph}}, \citenamefont {{Schlom}},
  \citenamefont {{{\'I}{\~n}iguez}}, \citenamefont {{Huey}},\ and\
  \citenamefont {{Ramesh}}}]{HER-14}%
  \BibitemOpen
  \bibfield  {author} {\bibinfo {author} {\bibfnamefont {J.~T.}\ \bibnamefont
  {{Heron}}}, \bibinfo {author} {\bibfnamefont {J.~L.}\ \bibnamefont
  {{Bosse}}}, \bibinfo {author} {\bibfnamefont {Q.}~\bibnamefont {{He}}},
  \bibinfo {author} {\bibfnamefont {Y.}~\bibnamefont {{Gao}}}, \bibinfo
  {author} {\bibfnamefont {M.}~\bibnamefont {{Trassin}}}, \bibinfo {author}
  {\bibfnamefont {L.}~\bibnamefont {{Ye}}}, \bibinfo {author} {\bibfnamefont
  {J.~D.}\ \bibnamefont {{Clarkson}}}, \bibinfo {author} {\bibfnamefont
  {C.}~\bibnamefont {{Wang}}}, \bibinfo {author} {\bibfnamefont
  {J.}~\bibnamefont {{Liu}}}, \bibinfo {author} {\bibfnamefont
  {S.}~\bibnamefont {{Salahuddin}}}, \bibinfo {author} {\bibfnamefont {D.~C.}\
  \bibnamefont {{Ralph}}}, \bibinfo {author} {\bibfnamefont {D.~G.}\
  \bibnamefont {{Schlom}}}, \bibinfo {author} {\bibfnamefont {J.}~\bibnamefont
  {{{\'I}{\~n}iguez}}}, \bibinfo {author} {\bibfnamefont {B.~D.}\ \bibnamefont
  {{Huey}}}, \ and\ \bibinfo {author} {\bibfnamefont {R.}~\bibnamefont
  {{Ramesh}}},\ }\href {\doibase 10.1038/nature14004} {\bibfield  {journal}
  {\bibinfo  {journal} {Nature}\ }\textbf {\bibinfo {volume} {516}},\ \bibinfo
  {pages} {370} (\bibinfo {year} {2014})}\BibitemShut {NoStop}%
\bibitem [{\citenamefont {Lahtinen}\ \emph {et~al.}(2011)\citenamefont
  {Lahtinen}, \citenamefont {Tuomi},\ and\ \citenamefont {van
  Dijken}}]{LAH-11}%
  \BibitemOpen
  \bibfield  {author} {\bibinfo {author} {\bibfnamefont {T.~H.~E.}\
  \bibnamefont {Lahtinen}}, \bibinfo {author} {\bibfnamefont {J.~O.}\
  \bibnamefont {Tuomi}}, \ and\ \bibinfo {author} {\bibfnamefont
  {S.}~\bibnamefont {van Dijken}},\ }\href {\doibase 10.1002/adma.201100426}
  {\bibfield  {journal} {\bibinfo  {journal} {Advanced Materials}\ }\textbf
  {\bibinfo {volume} {23}},\ \bibinfo {pages} {3187} (\bibinfo {year}
  {2011})}\BibitemShut {NoStop}%
\bibitem [{\citenamefont {{Lahtinen}}\ \emph
  {et~al.}(2012{\natexlab{a}})\citenamefont {{Lahtinen}}, \citenamefont
  {{Franke}},\ and\ \citenamefont {{van Dijken}}}]{LAH-12}%
  \BibitemOpen
  \bibfield  {author} {\bibinfo {author} {\bibfnamefont {T.~H.~E.}\
  \bibnamefont {{Lahtinen}}}, \bibinfo {author} {\bibfnamefont {K.~J.~A.}\
  \bibnamefont {{Franke}}}, \ and\ \bibinfo {author} {\bibfnamefont
  {S.}~\bibnamefont {{van Dijken}}},\ }\href@noop {} {\bibfield  {journal}
  {\bibinfo  {journal} {Scientific Reports}\ }\textbf {\bibinfo {volume} {2}},\
  \bibinfo {eid} {258} (\bibinfo {year} {2012}{\natexlab{a}})}\BibitemShut
  {NoStop}%
\bibitem [{\citenamefont {{Lahtinen}}\ \emph
  {et~al.}(2012{\natexlab{b}})\citenamefont {{Lahtinen}}, \citenamefont
  {{Shirahata}}, \citenamefont {{Yao}}, \citenamefont {{Franke}}, \citenamefont
  {{Venkataiah}}, \citenamefont {{Taniyama}},\ and\ \citenamefont {{van
  Dijken}}}]{LAH-12a}%
  \BibitemOpen
  \bibfield  {author} {\bibinfo {author} {\bibfnamefont {T.~H.~E.}\
  \bibnamefont {{Lahtinen}}}, \bibinfo {author} {\bibfnamefont
  {Y.}~\bibnamefont {{Shirahata}}}, \bibinfo {author} {\bibfnamefont
  {L.}~\bibnamefont {{Yao}}}, \bibinfo {author} {\bibfnamefont {K.~J.~A.}\
  \bibnamefont {{Franke}}}, \bibinfo {author} {\bibfnamefont {G.}~\bibnamefont
  {{Venkataiah}}}, \bibinfo {author} {\bibfnamefont {T.}~\bibnamefont
  {{Taniyama}}}, \ and\ \bibinfo {author} {\bibfnamefont {S.}~\bibnamefont
  {{van Dijken}}},\ }\href {\doibase 10.1063/1.4773482} {\bibfield  {journal}
  {\bibinfo  {journal} {Applied Physics Letters}\ }\textbf {\bibinfo {volume}
  {101}},\ \bibinfo {pages} {262405} (\bibinfo {year}
  {2012}{\natexlab{b}})}\BibitemShut {NoStop}%
\bibitem [{\citenamefont {{Chopdekar}}\ \emph {et~al.}(2012)\citenamefont
  {{Chopdekar}}, \citenamefont {{Malik}}, \citenamefont {{Fraile
  Rodr{\'{\i}}guez}}, \citenamefont {{Le Guyader}}, \citenamefont {{Takamura}},
  \citenamefont {{Scholl}}, \citenamefont {{Stender}}, \citenamefont
  {{Schneider}}, \citenamefont {{Bernhard}}, \citenamefont {{Nolting}},\ and\
  \citenamefont {{Heyderman}}}]{CHO-12}%
  \BibitemOpen
  \bibfield  {author} {\bibinfo {author} {\bibfnamefont {R.~V.}\ \bibnamefont
  {{Chopdekar}}}, \bibinfo {author} {\bibfnamefont {V.~K.}\ \bibnamefont
  {{Malik}}}, \bibinfo {author} {\bibfnamefont {A.}~\bibnamefont {{Fraile
  Rodr{\'{\i}}guez}}}, \bibinfo {author} {\bibfnamefont {L.}~\bibnamefont {{Le
  Guyader}}}, \bibinfo {author} {\bibfnamefont {Y.}~\bibnamefont {{Takamura}}},
  \bibinfo {author} {\bibfnamefont {A.}~\bibnamefont {{Scholl}}}, \bibinfo
  {author} {\bibfnamefont {D.}~\bibnamefont {{Stender}}}, \bibinfo {author}
  {\bibfnamefont {C.~W.}\ \bibnamefont {{Schneider}}}, \bibinfo {author}
  {\bibfnamefont {C.}~\bibnamefont {{Bernhard}}}, \bibinfo {author}
  {\bibfnamefont {F.}~\bibnamefont {{Nolting}}}, \ and\ \bibinfo {author}
  {\bibfnamefont {L.~J.}\ \bibnamefont {{Heyderman}}},\ }\href {\doibase
  10.1103/PhysRevB.86.014408} {\bibfield  {journal} {\bibinfo  {journal}
  {Physical Review B}\ }\textbf {\bibinfo {volume} {86}},\ \bibinfo {pages}
  {014408} (\bibinfo {year} {2012})}\BibitemShut {NoStop}%
\bibitem [{\citenamefont {{Streubel}}\ \emph {et~al.}(2013)\citenamefont
  {{Streubel}}, \citenamefont {{K{\"o}hler}}, \citenamefont {{Sch{\"a}fer}},\
  and\ \citenamefont {{Eng}}}]{STR-13}%
  \BibitemOpen
  \bibfield  {author} {\bibinfo {author} {\bibfnamefont {R.}~\bibnamefont
  {{Streubel}}}, \bibinfo {author} {\bibfnamefont {D.}~\bibnamefont
  {{K{\"o}hler}}}, \bibinfo {author} {\bibfnamefont {R.}~\bibnamefont
  {{Sch{\"a}fer}}}, \ and\ \bibinfo {author} {\bibfnamefont {L.~M.}\
  \bibnamefont {{Eng}}},\ }\href {\doibase 10.1103/PhysRevB.87.054410}
  {\bibfield  {journal} {\bibinfo  {journal} {Physical Review B}\ }\textbf
  {\bibinfo {volume} {87}},\ \bibinfo {pages} {054410} (\bibinfo {year}
  {2013})}\BibitemShut {NoStop}%
\bibitem [{\citenamefont {Ghidini}\ \emph {et~al.}(2015)\citenamefont
  {Ghidini}, \citenamefont {Maccherozzi}, \citenamefont {Moya}, \citenamefont
  {Phillips}, \citenamefont {Yan}, \citenamefont {Soussi}, \citenamefont
  {Métallier}, \citenamefont {Vickers}, \citenamefont {Steinke}, \citenamefont
  {Mansell}, \citenamefont {Barnes}, \citenamefont {Dhesi},\ and\ \citenamefont
  {Mathur}}]{GHI-15}%
  \BibitemOpen
  \bibfield  {author} {\bibinfo {author} {\bibfnamefont {M.}~\bibnamefont
  {Ghidini}}, \bibinfo {author} {\bibfnamefont {F.}~\bibnamefont
  {Maccherozzi}}, \bibinfo {author} {\bibfnamefont {X.}~\bibnamefont {Moya}},
  \bibinfo {author} {\bibfnamefont {L.~C.}\ \bibnamefont {Phillips}}, \bibinfo
  {author} {\bibfnamefont {W.}~\bibnamefont {Yan}}, \bibinfo {author}
  {\bibfnamefont {J.}~\bibnamefont {Soussi}}, \bibinfo {author} {\bibfnamefont
  {N.}~\bibnamefont {Métallier}}, \bibinfo {author} {\bibfnamefont {M.~E.}\
  \bibnamefont {Vickers}}, \bibinfo {author} {\bibfnamefont {N.~J.}\
  \bibnamefont {Steinke}}, \bibinfo {author} {\bibfnamefont {R.}~\bibnamefont
  {Mansell}}, \bibinfo {author} {\bibfnamefont {C.~H.~W.}\ \bibnamefont
  {Barnes}}, \bibinfo {author} {\bibfnamefont {S.~S.}\ \bibnamefont {Dhesi}}, \
  and\ \bibinfo {author} {\bibfnamefont {N.~D.}\ \bibnamefont {Mathur}},\
  }\href {\doibase 10.1002/adma.201404799} {\bibfield  {journal} {\bibinfo
  {journal} {Advanced Materials}\ }\textbf {\bibinfo {volume} {27}},\ \bibinfo
  {pages} {1460} (\bibinfo {year} {2015})}\BibitemShut {NoStop}%
\bibitem [{\citenamefont {{Franke}}\ \emph {et~al.}(2015)\citenamefont
  {{Franke}}, \citenamefont {{Van de Wiele}}, \citenamefont {{Shirahata}},
  \citenamefont {{H{\"a}m{\"a}l{\"a}inen}}, \citenamefont {{Taniyama}},\ and\
  \citenamefont {{van Dijken}}}]{FRA-15}%
  \BibitemOpen
  \bibfield  {author} {\bibinfo {author} {\bibfnamefont {K.~J.~A.}\
  \bibnamefont {{Franke}}}, \bibinfo {author} {\bibfnamefont {B.}~\bibnamefont
  {{Van de Wiele}}}, \bibinfo {author} {\bibfnamefont {Y.}~\bibnamefont
  {{Shirahata}}}, \bibinfo {author} {\bibfnamefont {S.~J.}\ \bibnamefont
  {{H{\"a}m{\"a}l{\"a}inen}}}, \bibinfo {author} {\bibfnamefont
  {T.}~\bibnamefont {{Taniyama}}}, \ and\ \bibinfo {author} {\bibfnamefont
  {S.}~\bibnamefont {{van Dijken}}},\ }\href {\doibase
  10.1103/PhysRevX.5.011010} {\bibfield  {journal} {\bibinfo  {journal}
  {Physical Review X}\ }\textbf {\bibinfo {volume} {5}},\ \bibinfo {eid}
  {011010} (\bibinfo {year} {2015})}\BibitemShut {NoStop}%
\bibitem [{\citenamefont {{Shirahata}}\ \emph {et~al.}(2015)\citenamefont
  {{Shirahata}}, \citenamefont {{Shiina}}, \citenamefont {{L{\'o}pez
  Gonz{\'a}lez}}, \citenamefont {{Franke}}, \citenamefont {{Wada}},
  \citenamefont {{Itoh}}, \citenamefont {{Pertsev}}, \citenamefont {{van
  Dijken}},\ and\ \citenamefont {{Taniyama}}}]{SHI-15}%
  \BibitemOpen
  \bibfield  {author} {\bibinfo {author} {\bibfnamefont {Y.}~\bibnamefont
  {{Shirahata}}}, \bibinfo {author} {\bibfnamefont {R.}~\bibnamefont
  {{Shiina}}}, \bibinfo {author} {\bibfnamefont {D.}~\bibnamefont {{L{\'o}pez
  Gonz{\'a}lez}}}, \bibinfo {author} {\bibfnamefont {K.~J.~A.}\ \bibnamefont
  {{Franke}}}, \bibinfo {author} {\bibfnamefont {E.}~\bibnamefont {{Wada}}},
  \bibinfo {author} {\bibfnamefont {M.}~\bibnamefont {{Itoh}}}, \bibinfo
  {author} {\bibfnamefont {N.~A.}\ \bibnamefont {{Pertsev}}}, \bibinfo {author}
  {\bibfnamefont {S.}~\bibnamefont {{van Dijken}}}, \ and\ \bibinfo {author}
  {\bibfnamefont {T.}~\bibnamefont {{Taniyama}}},\ }\href {\doibase
  10.1038/am.2015.72} {\bibfield  {journal} {\bibinfo  {journal} {NPG Asia
  Materials}\ }\textbf {\bibinfo {volume} {7}},\ \bibinfo {pages} {e198}
  (\bibinfo {year} {2015})}\BibitemShut {NoStop}%
\bibitem [{\citenamefont {{Franke}}\ \emph {et~al.}(2012)\citenamefont
  {{Franke}}, \citenamefont {{Lahtinen}},\ and\ \citenamefont {{van
  Dijken}}}]{FRA-12}%
  \BibitemOpen
  \bibfield  {author} {\bibinfo {author} {\bibfnamefont {K.~J.~A.}\
  \bibnamefont {{Franke}}}, \bibinfo {author} {\bibfnamefont {T.~H.~E.}\
  \bibnamefont {{Lahtinen}}}, \ and\ \bibinfo {author} {\bibfnamefont
  {S.}~\bibnamefont {{van Dijken}}},\ }\href {\doibase
  10.1103/PhysRevB.85.094423} {\bibfield  {journal} {\bibinfo  {journal}
  {Physical Review B}\ }\textbf {\bibinfo {volume} {85}},\ \bibinfo {pages}
  {094423} (\bibinfo {year} {2012})}\BibitemShut {NoStop}%
\bibitem [{\citenamefont {{Franke}}\ \emph {et~al.}(2014)\citenamefont
  {{Franke}}, \citenamefont {{L{\'o}pez Gonz{\'a}lez}}, \citenamefont
  {{H{\"a}m{\"a}l{\"a}inen}},\ and\ \citenamefont {{van Dijken}}}]{FRA-14}%
  \BibitemOpen
  \bibfield  {author} {\bibinfo {author} {\bibfnamefont {K.~J.~A.}\
  \bibnamefont {{Franke}}}, \bibinfo {author} {\bibfnamefont {D.}~\bibnamefont
  {{L{\'o}pez Gonz{\'a}lez}}}, \bibinfo {author} {\bibfnamefont {S.~J.}\
  \bibnamefont {{H{\"a}m{\"a}l{\"a}inen}}}, \ and\ \bibinfo {author}
  {\bibfnamefont {S.}~\bibnamefont {{van Dijken}}},\ }\href {\doibase
  10.1103/PhysRevLett.112.017201} {\bibfield  {journal} {\bibinfo  {journal}
  {Physical Review Letters}\ }\textbf {\bibinfo {volume} {112}},\ \bibinfo
  {pages} {017201} (\bibinfo {year} {2014})}\BibitemShut {NoStop}%
\bibitem [{\citenamefont {{Casiraghi}}\ \emph {et~al.}(2015)\citenamefont
  {{Casiraghi}}, \citenamefont {{Rinc{\'o}n Dom{\'{\i}}nguez}}, \citenamefont
  {{R{\"o}{\ss}ler}}, \citenamefont {{Franke}}, \citenamefont {{L{\'o}pez
  Gonz{\'a}lez}}, \citenamefont {{H{\"a}m{\"a}l{\"a}inen}}, \citenamefont
  {{Fr{\"o}mter}}, \citenamefont {{Oepen}},\ and\ \citenamefont {{van
  Dijken}}}]{CAS-15}%
  \BibitemOpen
  \bibfield  {author} {\bibinfo {author} {\bibfnamefont {A.}~\bibnamefont
  {{Casiraghi}}}, \bibinfo {author} {\bibfnamefont {T.}~\bibnamefont
  {{Rinc{\'o}n Dom{\'{\i}}nguez}}}, \bibinfo {author} {\bibfnamefont
  {S.}~\bibnamefont {{R{\"o}{\ss}ler}}}, \bibinfo {author} {\bibfnamefont
  {K.~J.~A.}\ \bibnamefont {{Franke}}}, \bibinfo {author} {\bibfnamefont
  {D.}~\bibnamefont {{L{\'o}pez Gonz{\'a}lez}}}, \bibinfo {author}
  {\bibfnamefont {S.~J.}\ \bibnamefont {{H{\"a}m{\"a}l{\"a}inen}}}, \bibinfo
  {author} {\bibfnamefont {R.}~\bibnamefont {{Fr{\"o}mter}}}, \bibinfo {author}
  {\bibfnamefont {H.~P.}\ \bibnamefont {{Oepen}}}, \ and\ \bibinfo {author}
  {\bibfnamefont {S.}~\bibnamefont {{van Dijken}}},\ }\href {\doibase
  10.1103/PhysRevB.92.054406} {\bibfield  {journal} {\bibinfo  {journal}
  {Physical Review B}\ }\textbf {\bibinfo {volume} {92}},\ \bibinfo {pages}
  {054406} (\bibinfo {year} {2015})}\BibitemShut {NoStop}%
\bibitem [{\citenamefont {{Vansteenkiste}}\ \emph {et~al.}(2014)\citenamefont
  {{Vansteenkiste}}, \citenamefont {{Leliaert}}, \citenamefont {{Dvornik}},
  \citenamefont {{Helsen}}, \citenamefont {{Garcia-Sanchez}},\ and\
  \citenamefont {{Van Waeyenberge}}}]{VAN-14}%
  \BibitemOpen
  \bibfield  {author} {\bibinfo {author} {\bibfnamefont {A.}~\bibnamefont
  {{Vansteenkiste}}}, \bibinfo {author} {\bibfnamefont {J.}~\bibnamefont
  {{Leliaert}}}, \bibinfo {author} {\bibfnamefont {M.}~\bibnamefont
  {{Dvornik}}}, \bibinfo {author} {\bibfnamefont {M.}~\bibnamefont {{Helsen}}},
  \bibinfo {author} {\bibfnamefont {F.}~\bibnamefont {{Garcia-Sanchez}}}, \
  and\ \bibinfo {author} {\bibfnamefont {B.}~\bibnamefont {{Van
  Waeyenberge}}},\ }\href {\doibase 10.1063/1.4899186} {\bibfield  {journal}
  {\bibinfo  {journal} {AIP Advances}\ }\textbf {\bibinfo {volume} {4}},\
  \bibinfo {eid} {107133} (\bibinfo {year} {2014})}\BibitemShut {NoStop}%
\bibitem [{\citenamefont {{Hubert}}(1979)}]{HUB-79}%
  \BibitemOpen
  \bibfield  {author} {\bibinfo {author} {\bibfnamefont {A.}~\bibnamefont
  {{Hubert}}},\ }\href {\doibase 10.1109/TMAG.1979.1060325} {\bibfield
  {journal} {\bibinfo  {journal} {IEEE Trans. Magn.}\ }\textbf {\bibinfo
  {volume} {15}},\ \bibinfo {pages} {1251} (\bibinfo {year}
  {1979})}\BibitemShut {NoStop}%
\end{thebibliography}


%

\end{document}